\documentclass[%
 reprint,
 amsmath,amssymb,
 aps,
]{revtex4-1}

\usepackage{color}
\def\ph{\color{black}}
\def\gi{\color{black}}
\def\phh{\color{black}}
\def\gii{\color{black}}
\usepackage{graphicx}
\usepackage{dcolumn}
\usepackage{bm}

\begin{document}

\preprint{APS/123-QED}

\title{ Nuclear symmetry energy and the role of three-body forces 
}

\author{S. Goudarzi}

\author{H. R. Moshfegh}%
\affiliation{%
 Department of Physics, University of Tehran, Post Office Box 14395-547, Tehran, Iran 
}%
\author{P. Haensel}%
\affiliation{%
N. Copernicus Astronomical Center, Polish Academy of Science, Bartycka 18, 00-716, Warszawa, Poland
}%

\date{\today}

\begin{abstract}

Density dependence of nuclear symmetry energy as well as its partial wave decomposition is studied within the framework of lowest-order constrained variational (LOCV) method using AV18  two-body interaction supplemented by UIX  {\ph{ three-body force}}. The main focus of the present work is to introduce a revised version of {\ph three-body force} which is based on an isospin-dependent parametrization of  {\ph coefficients}  in the UIX force, in order to overcome the inability to produce correct {\phh saturation-point parameters}  in the framework of LOCV method.  We find that employing the new model of {\ph three-body force} in the LOCV formalism leads to  successfully reproducing  the {\ph semi-empirical parameters} of cold nuclear matter, {\phh including}  $E_{sym}(\rho_0)$, $L$, and $K_{sym}$. {\phh{All our  models of three-body force  combined with  AV18 
two-body force give  maximum neutron star mass higher than  $2\;M_\odot$. The fraction of protons in the nucleon cores of neutron stars strongly depends on the three-body force parametrization. }} 

\begin{description}
\item[PACS numbers]
21.30.Fe,21.65.Mn,21.65.Cd,21.65.Ef
\end{description}
\end{abstract}

\maketitle


\section{Introduction}

One of the aims of nuclear science is to understand the properties of strongly interacting bulk matter at nuclear levels. In this regard, properties of symmetric nuclear matter (SNM) has been studied for a long time and its equation of state (EOS) around saturation density $\rho_{0}$ is known rather well. On the other hand, the EOS of isospin asymmetric nuclear matter, particularly pure neutron matter (PNM), has not been established well despite of its crucial role in both nuclear physics and astrophysics. Motivated by this fact, studying the nuclear symmetry energy $E_{sym}$ which mostly governs the PNM EOS is currently an active field of research in nuclear physics.

 Since the symmetry energy  can not be directly measured, it is of fundamental importance to identify observables which strongly correlate  with $E_{sym}$ and its density slope $L$ to impose constraints on these quantities. Additional information on the value and density dependence of symmetry energy can be extracted from the astrophysical observations of compact objects, namely neutron stars. Among the terrestrial laboratory observables, the neutron skin thickness is widely studied \cite{1,2,3,4,5,6} during last decades. Nowadays, there are many other observables such as isospin fractionation, neutron-proton differential flow, nuclear mass systematics, low-lying E1 mode, etc., which are believed to be sensitive to symmetry energy and particularly, its density slope. We refer to references {\ph \cite{7,8,9,10,11}} and references therein quoted for more details. Density dependence of this quantity also relates the heavy-ion reactions \cite{12,13,14,15,16,17,18}, stability of superheavy nuclei \cite{19}, cooling \cite{20,21,22,23,24} and the mass-radius relations of neutron stars \cite{25,26}, and properties of nuclei involved in r-process nucleosynthesis \cite{27}.
 
 In order to theoretically predict the symmetry energy, asymmetric nuclear matter is studied in the framework of various microscopic and phenomenological approaches by using a variety of microscopic and phenomenological two-nucleon forces and phenomenological three-body forces. Such many-body techniques include the {\ph Brueckner-Hartree-Fock} (BHF) \cite{7,28}, its relativistic counterpart Dirac-Brueckner-Hartree-Fock (DBHF) \cite{29}, the self-consistent {\ph Green's} function \cite{30}, variational approaches \cite{31,32}, mean-field model \cite{33}, the Quantum Monte Carlo (QMC) \cite{34}, extended relativistic mean field (ERMF) model \cite{35}, and Skyrme-Hartree-Fock (SHF) method \cite{36}. There are also some recent calculations for the symmetry energy of finite nuclei (see e.g. \cite{37} and references therein). Recent progress and challenges in both theoretically and experimentally measuring  the symmetry energy and its density dependence, specially at supra nuclear densities, are also reviewed in Refs. \cite{38,39,40,41}.
 
 Despite of numerous attempts that was made to determine the symmetry energy properties, especially the value of $E_{sym}(\rho_{0})$ and its density slope  $L$ at saturation density $\rho_{0}$, these quantities are still uncertain. Results are strongly dependent on different experimental techniques and many-body approaches as well as nuclear interactions. Especially, the density dependence of symmetry energy is poorly known. The high density prediction of many-body models for the symmetry energy are extremely diverse and sometimes contradictory. On the other hand, although the value of the symmetry energy is known to be in the range of $\sim 31.6\pm 2.66 $ MeV \cite{40}, explorations on $L$ show wide variations. The obtained lower limit is $\sim 20 $ MeV \cite{42} and the upper one is even higher than 170 MeV \cite{43}. However, recent surveys of analyses of terrestrial nuclear laboratory experiments and astrophysical observations found that the mean value of the slope parameter at saturation density lies in the range of $ 58.9\pm 16.0 $ MeV {\ph \cite{40,11}}.
 
 In the present work, within the self-consistent lowest order constrained variational (LOCV) approach using the {\phh Argonne} V18 two-body potential \cite{44} supplemented by an Urbana type three-body force (UIX) \cite{45}, we study the density dependence of nuclear symmetry energy with the focus on the effect of three-body interaction  on this quantity. The LOCV method is a well-known many-body technique that was originally used to study the properties of cold symmetric nuclear matter  \cite{46,47}  by using the Reid-type potential \cite{48,49}  as the bare two-body interaction. Later on, this approach was extended to finite temperature \cite{50} and also calculation of the EOS of asymmetric nuclear matter  \cite{51},  pure neutron matter \cite{52},  and beta- stable matter \cite{53}  was carried out within this framework by using more sophisticated potentials. Moreover, relativistic corrections has been considered in calculating thermodynamic properties of nuclear matter within this model at both zero \cite{54}  and finite temperature \cite{55}. Recently, this technique has been extended by adding three-body forces to its formalism \cite{56} and has been used to study the structure of neutron stars \cite{56}  as well as proto-neutron stars  \cite{57}.
 
  Although this method was successful in reproducing some properties that characterize the cold symmetric nuclear matter, namely the saturation density, saturation energy and the incompressibility, the nuclear symmetry energy obtained by this model turns out to be high and does not lie in the proper range \cite{56}. Therefore, our goal in the present work is to correct this deficiency by revising the parametrization of UIX three-body interaction.

The article is organized as follows. In Sec. II we 
 review briefly the adopted LOCV many-body approach including three-body interaction.  Section III is devoted to theoretically studying the nuclear symmetry energy. 
 Results and discussion are presented in Sec. IV. Finally {\ph summary} and conclusion are given in Sec. V.

\section{Theoretical Formalism}

In this section we briefly describe the LOCV formalism whose details can be found in mentioned references in the introduction section.
As the first step of calculating the EOS of asymmetric nuclear matter within the LOCV approach, we produce a trial wave function for A-body interacting system at zero temperature, denoted by $\Psi$, which is defined as the product of a non-interacting ground state wave function of $A$ independent nucleons, $\Phi$,  and an $A$-body correlation operator $F$, as follows
 \cite{53}
\begin{equation}
\Psi(1...A)=F(1...A)\Phi(1...A).
\end{equation}
The correlation operator is in general considered  as the symmetrized product of two-body correlation function operators, i.e.,
\begin{equation}
F=\mathcal{S}\prod _{i>j}f(ij) ,
\end{equation}
where $\mathcal{S}$ is the symmetrizing operator. $f(ij)$ is expressed as
\begin{equation}
f(ij)=\sum _{\alpha ,p=1}^{3} f_{\alpha }^{(p)} (ij )O_{\alpha }^{(p)} (ij).
\end{equation}
Here, $\alpha$ stands for $J,L,S,T,$ and $T_{z}$.  $p$ is set to unity except for  triplet channels with $J=L\pm1$. In this case we choose $p$=2 and 3 with $O_{\alpha }^{(2)}=\frac{2}{3} +\frac{1}{6} S_{12}$  and $O_{\alpha }^{(3)}=\frac{1}{3} -\frac{1}{6} S_{12}$.
The operator $S_{12}$  denotes the usual tensor operator. 
In general, the nuclear Hamiltonian includes a non-relativistic one-body kinetic energy as well as a two-body potential $V(ij)$,
\begin{equation}
H=\sum _{i}\frac{p_{i} ^{2} }{2m_{i}}  +\sum _{i<j}V(ij).
\end{equation}
By neglecting the small contribution from higher-order terms in the cluster expansion series \cite{50}, the energy expectation value $E$ is gained via
\begin{equation}
E[f]=\frac{1}{A} \frac{{\left\langle \Psi  \right|} H{\left| \Psi  \right\rangle} }{{\left\langle \Psi  \mathrel{\left| \vphantom{\Psi  \Psi }\right.\kern-\nulldelimiterspace} \Psi  \right\rangle} } =E_{1} +E_{MB} \cong E_{1} +E_{2},
\end{equation}
with
\begin{equation}
 E_{1} =\sum _{i}\frac{3\hbar ^{2} {k^F_{i}}^{2} }{10m_{i}},
\end{equation}
 where  $i=n,p$, $k^F_{i}$ is corresponding nucleon  Fermi momentum divided by $\hbar$, 
and 
\begin{equation}
E_{2} =\frac{1}{2A} \sum _{ij}{\left\langle ij \right|}  W(12){\left| ij-ji \right\rangle},
\end{equation}
where $W(12)$ is expressed as
\begin{equation}
W(12)=-\frac{\hbar ^{2} }{2m} [f(12),[\nabla _{12}^{2} ,f(12)]]+f(12)V(12)f(12). 
\end{equation}
After doing some algebra, final expression for the two-body cluster energy is 
obtained  by minimizing expression given by Eq.(8) with respect to variations in the functions $f_{\alpha }^{(p)}$ under the condition
\begin{equation}
\frac{1}{A} \sum _{ij}\left\langle ij|h_{T_{z} }^{2} (12)-f_{}^{2} (12)|ij-ji\right\rangle  =0.
\end{equation}
Equation (9) defines the LOCV normalisation constraint and can be calculated by using the modified Pauli function $h_{T_{z}}(r)$ which in case of asymmetric nuclear matter takes the following form
\begin{eqnarray}
{\ph h_{T_{z} }(r)}&&={\ph \left[
1-{\frac{9}{2}} \left(
{{J_{1}\left(k_{i}^{ F} r\right)}\over {k_{i}^{F} r}} 
\right)^{2} 
\right]^{-\frac{1}{2}}}~~~~~~,{\ph ~~T_{z}} =\pm 1
\nonumber\\
&&=1 ~~~~~~,{\ph ~~T_{z} }=0.
\end{eqnarray}
$J_{L} (x)$ denotes the spherical Bessel function of order $L$. The normalisation constraint introduces another parameter in the LOCV formalism, i.e. the Lagrange multiplier $\lambda$. Procedure of minimizing  Eq. (8) leads to a number of Euler-Lagrange differential equations for functions $ f^{(p)}_\alpha(ij)$. By solving these equations, correlation functions and consequently, the two-body cluster energy can be determined.

The important role played by three-body forces {\ph (3BF)} in both finite nuclei and nuclear matter calculations is well known. In nuclear matter calculations, unfortunately, whatever realistic two-body force {\ph (2BF)} is used in a many-body approach, the saturation properties of cold symmetric nuclear matter fail to be reproduced correctly. This deficiency can be cured by inclusion of a {\ph 3BF}  in the nuclear Hamiltonian. For this reason, the semiphenomenological UIX interaction is considered in our many-body calculations. Generally, the UIX interaction is written as following \cite{45}
\begin{equation}
V_{123} =V_{123}^{2\pi } +V_{123}^{R},
\end{equation}
with a two-pion exchange contribution as
\begin{eqnarray}
V_{123}^{2\pi } =A\sum _{cyc}(\{ X_{12} , X_{23} \} \{ {\ph \vec{\tau}_{1} \cdot\vec{\tau}_{2} ,
\vec{\tau}_{2}\cdot\vec{\tau}_{3}} \} 
\nonumber\\
+\frac{1}{4} [X_{12} ,X_{23} ]{\ph [ \vec{\tau}_{1}\cdot \vec{\tau}_{2},\vec{\tau}_{2} \cdot \vec{\tau}_{3} ]}), 
\end{eqnarray}
and a shorter-range phenomenological part,
\begin{equation}
V_{123}^{R} =U\sum _{cyc}T^{2}(m_{\pi } r_{12}) T^{2}(m_{\pi } r_{23} )
\end{equation}
$A$ and $U$ are adjustable parameters determined by fitting the empirical saturation density and energy
of the cold SNM in the LOCV calculations. Numerals 1, 2, and 3 stand for three different interacting nucleons while $\sigma$, $\tau$, and $S_{12}$ denotes the  spin, isospin and the usual tensor operator respectively and $Y(m_{\pi } r)$ and $T(m_{\pi } r)$ are the Yukawa and tensor functions. The one-pion exchange operator $X_{12}$ in Eq. (12) is defined as
\begin{equation}
X_{12} =Y(m_{\pi } r_{12} ){\ph \vec{\sigma} _{1} \cdot \vec{\sigma} _{2}} +T(m_{\pi } r_{12} )S_{12}.
\end{equation}
In two recent papers we have presented the procedure of extending the LOCV approach by using this kind of {\ph 3BF} in its formalism at both zero \cite{56} and finite temperature \cite{57}. The  {\ph 3BF} is included via an effective two-body potential derived after averaging out the third particle, being weighted by the LOCV two-body correlation functions at fixed density $\rho$, i.e.,
\begin{equation}
\bar{V}_{12}^{} (r,T)=\rho \int d^{3} r_{3}  \sum _{\sigma _{3} ,\tau _{3} }f^{2} (r_{13} )f^{2} (r_{23} )V_{123},
\end{equation}
 where  $r=r_{12}$.  
 In this way, a density-dependent effective two-body interaction is gained with the following operator structure
\begin{eqnarray}
\bar{V}_{12}^{} (r)=&&{\ph (\vec{\tau} _{1} \cdot \vec{\tau} _{2} )(\vec{\sigma} _{1}\cdot \vec{\sigma} _{2}} )V^{{2\pi}}_{\sigma \tau } (r)
\nonumber\\
&&+S_{12} (\hat{r})({\ph \vec{\tau}_{1} \cdot \vec{\tau}_{2}} )V^{{2\pi}}_{t}(r)
+V^{{R}}_{c}(r).
\end{eqnarray}
Explicit relations for the coefficients $V^{{2\pi}}_{\sigma \tau } (r)$, $V^{{2\pi}}_{t}(r)$, and $V^{{R}}_{c}(r)$ are expressed in \cite{56}. It is pointed out in the mentioned reference that some quantities that characterizes the saturation point of cold symmetric nuclear matter can not be produced within the LOCV approach even if the nuclear Hamiltonian is supplemented by a {\ph 3BF}. This deficiency motivated us to propose a new parametrization for UIX {\ph 3BF} which is based on isospin-dependent strengths. Our main purpose in the present work is to construct such {\ph a 3BF}  in order to reach correct saturation properties within the LOCV model. 
In this regard, {\ph subsection B of Sect.IV } is devoted to investigating this issue in detail. {\ph The next Section III refers to the nuclear symmetry energy and other parameters characterizing saturation point of nuclear matter. }

\begin{table*}
\caption{\label{tab:table2}%
Saturation properties of symmetric nuclear matter for LOCV calculations with and without {\ph 3BF} and several many-body techniques and interactions. All the quantities are in MeV with the exception of $\rho_{0}$ given in {\ph ${\rm fm^{-3}}$}.
}
\begin{ruledtabular}
\begin{tabular}{lccccccc}
\textrm{Model}&
\textrm{$\rho_{0}$}&
\textrm{$E_{0}/A$}&
\textrm{$K_{0}$ }&
\textrm{$E_{sym}(\rho_0)$ }&
\textrm{$L $}&
\textrm{$K_{sym}$}&
\textrm{$K_{asy}$}\\
\colrule
LOCV (AV18) \cite{56} & 0.3270 & -23.37 & 373.31 & 40.52 & 74.07 & -76.30 & -520.57   \\

BHF (AV18) \cite{28} & 0.240 & -17.30 & 213.6  & 35.8 & 63.1 & -27.8  &  -339.6   \\

QMC (AV8) \cite{34} & 0.16 & -16.0 & -  & 30.50 & 31.30 & -  &  -   \\

DBHF (Bonn A) \cite{29} & 0.181 & -16.62 & 233  & 34.80 & 71.20 & -  &  -   \\
DBHF (Bonn B) \cite{29} & 0.162 & -15.04 & 190  & 31.20 & 55.90 & -  &  -   \\
DBHF (Bonn C) \cite{29} & 0.148 & -14.14 & 170  & 28.9 & 46.7 & -  &  -   \\
DBHF (Bonn AB) \cite{29} & 0.17 & -15.52 & 204  & 32.3 & 61.1 & -  &  -   \\

LOCV (AV18+UIX) \cite{56} & 0.1748 & -15.58 & 295.77 & 39.9 & 115.75 & 15.58 & -678.92   \\

BHF(AV18+UIX) \cite{7} & 0.187 & -15.23 & -  & 34.30 & 66.50 & -  &  -   \\

QMC (AV8+UIX) \cite{34} & 0.16 & -16.0 & -  & 35.10 & 63.60 & -  &  -   \\

variational (AV8+UIX) \cite{32} & 0.16 & -16.09 & 245  & 30.0 & -  & -  &   -  \\

BHF (AV18+UIX) \cite{28} & 0.176 & -14.62 & 185.9  & 33.6 & 66.9 & -23.4  &  -343.8   \\

mean field \cite{33} & 0.16 & -16.0  & -  & $31.6 \pm 2.2$ & $56 \pm 24$ & -125.79  &  -   \\

ERMF (BSR1) \cite{35} & 0.1481 & -16.02 & 240.05  & 30.98 & 59.61 & -  &  -   \\
ERMF (BSR7) \cite{35} & 0.1493 & -16.17 & 231.86  & 36.99 & 98.78 & -  &  -   \\

SHF \cite{36} & 0.16 & -16.0 & 230.0  & $30.5 \pm 3$ & $52.5 \pm 20$ & -  &  -   \\

FSU \cite{61} & 0.148 & -16.30 & 230.0  & 32.6 & 60.5 & -51.3  &  -276.6   \\

NL3 \cite{62} & 0.148 & -16.24 & 271.6  & 37.4 & 118.5 & 100.9  &  -698.4   \\

TM1 \cite{63} & 0.145 & -16.32 & 281.0  & 36.8 & 110.8 & -66.4  &  -518.7   \\

DDME1 \cite{64} & 0.152 & -16.23 & 332.8  & 33.1 & 55.6 & -100.8  &  -508.1   \\

RMF (NL3) \cite{37} & 0.148 & -16.3 & 272  & 37.4 & 118.2 & -  &  -   \\
RMF (TM1) \cite{37} & 0.145 & -16.3 & 281  & 36.9 & 110.8 & -  &  -   \\
RMF (FSU) \cite{37} & 0.148 & -16.3 & 230  & 32.6 & 60.5 & -  &  -   \\
RMF (IUFSU) \cite{37} & 0.155 & -16.4 & 231  & 31.3 & 47.2 & -  &  -   \\

\end{tabular}
\end{ruledtabular}
\end{table*}

\section{Nuclear symmetry energy}
The energy per particle of cold asymmetric nuclear matter at given baryon density  $\rho(=\rho_n+\rho_p)$ and asymmetry parameter $X$($=(\rho_{n}-\rho_{p})/\rho$) can be expanded around $X=0$ as
\begin{equation}
E(\rho,X)=E(\rho,X=0)+E_{sym}(\rho)X^2+O(X^4).
\end{equation}
The expansion coefficient is the nuclear symmetry energy and is defined as following
\begin{equation}
E_{sym}^{exact}(\rho)=\dfrac{1}{2}\dfrac{\partial^2 E(\rho,X)}{\partial X^2}\mid_{X=0}.
\end{equation}
Due to the charge symmetry of nuclear forces, odd powers of $X$ are absent in Eq. (17). Magnitude of the $X^4$ term at saturation density has been estimated to be less than 1 MeV \cite{58}. Therefore,  higher-order terms in $X$ can be neglected compared to the value of the quadratic term. 
Equation  (17) is known as the parabolic approximation for the EOS of asymmetric nuclear matter which is accurate close to $X=0$ and also works well for higher values of the asymmetry parameter. {\ph Within} this approximation, the symmetry energy can be defined as
\begin{equation}
E_{sym}(\rho)=E(\rho,X=1)-E(\rho,X=0),
\end{equation}
which states that to a good approximation, the symmetry energy
measures the difference between the energy per particle
in {\ph uniform}  isospin-symmetric matter and pure neutron matter at the fixed density $\rho$. {\ph Then}, one can expand the nuclear symmetry energy around
the nuclear matter saturation density $\rho_0$. To second order the expansion is written as
\begin{equation}
{\ph E_{sym}(\rho)=E_{sym}(\rho_0)+\dfrac{L}{3}\left(\dfrac{\rho-\rho_0}{\rho_0}\right)+\dfrac{K_{sym}}{18}\left(\dfrac{\rho-\rho_0}{\rho_0}\right)^2,}
\end{equation}
where the coefficients respectively denote the nuclear symmetry energy value at normal density, the slope, and the curvature parameter. The $L$ and $K_{sym}$ characterize the density dependence of symmetry energy around the saturation density and are written as
\begin{equation}
L=3\rho_0\left({\phh{{{\rm d}E_{sym}(\rho)}\over {{\rm d}} \rho}}\right)_{\rho_0}.
\end{equation}
\begin{equation}
K_{sym}=9{\rho_0}^2\left({\phh {\dfrac{{\rm d}^2 E_{sym}(\rho)}{{\rm d} \rho^2}}}\right)_{\rho_0}.
\end{equation}
On the other hand, it is possible to expand the isobaric incompressibility of asymmetric nuclear matter in the power 
series of asymmetry parameter $X$ as
\begin{equation}
K(X)=K_0(X=0)+K_{asy}X^2+K_{4}X^4+O(X^6),
\end{equation}
where $K_0(X=0)$ is the incompressibility of SNM at normal density and is expressed as
\begin{equation}
K_0(X=0)=9{\rho_0}^2{\phh\left(\dfrac{{\rm d}^2 E(\rho,X=0)}{{\rm d} \rho^2}\right)_{\rho_0}}~.
\end{equation}
The $K_{asy}$ coefficient characterizes the isospin dependence of the incompressibility at saturation density and is often determined by using the following approximate expression \cite{59}
\begin{equation} 
K_{asy}\approx K_{sym}-6L~.
\end{equation}
Once again, one can keep only the first two terms in Eq. (23) and neglect higher order ones in $X$ \cite{60}.
 Quantities $K_{asy}$ and $E_{sym}(\rho_0)$ together with the slope parameter can be determined experimentally and provide information about the density dependence of {\ph nuclear symmetry energy} at normal nuclear density.

\section{RESULTS AND DISCUSSION}
We first focus on  calculations which are done by using only 2BF as well as {\ph 2BF+3BF} with isospin-independent parameters in UIX {\ph 3BF}. After that, in the second subsection we present our results {\ph obtained using }  revised version of {\ph 3BF} {\ph based on the} isospin-dependent parametrization of {\ph coefficients}  $A$ and $U$ in the UIX three-body force.

\subsection{UIX three-body force with isospin-independent A and U parameters}
In the beginning, we mention that in order to calculate the effective two-body potential via Eq. (15), the $^1S_0$ channel two-body correlation function is used, as discussed in our previous work \cite{56}. Moreover, for each specific asymmetry, the two-body correlation function which is obtained at the same asymmetry is used.
Let us now start with presenting results of the energy per nucleon. Figure 1 shows the comparison between the energy per particle of SNM $(X=0)$ and PNM $(X=1)$ using 2BF as well as 2BF+{\ph 3BF}. Properties of saturation point of SNM in both cases are also listed in Table I.  As has been mentioned earlier and is seen in {\ph Fig.1}, if two-body potential is the only interaction considered in the nuclear Hamiltonian, calculations  fail to reproduce the saturation point
of symmetric nuclear matter, i.e. they yield too {\ph small} binding energy and a saturation density well above the empirical value of 
$\rho_{0}=0.17\pm0.01 \;{\ph \rm fm^{-3}}$. Moreover, from Table I it can be concluded that the isobaric incompressibility of SNM, $K_0$, which is defined by Eq. (24), is too large and does not  lie in the rage of $250<K_0<315$ MeV which is calculated by re-analysing of recent data on the giant monopole resonance \cite{65}

 Now let us focus on the case of using both 2BF and {\ph 3BF} in the calculations. As expected, inclusion of {\ph 3BF} leads to an acceptable saturation point by increasing the {\ph binding energy at saturation} as well as pushing the saturation density toward {\ph lower values}. The isobaric incompressibility at normal density is also decreased so that it fits well in the proper range. It is found that values $A=-0.041$ MeV and $U=-0.000523$ MeV yield a reasonable saturation point for cold SNM in the LOCV calculations.

By using the empirical parabolic law, nuclear symmetry energy can be easily extracted from LOCV microscopic calculations via Eq. (19) provided the parabolic approximation is valid. Quadratic dependence of the symmetry energy {\ph $E(\rho,X)-E(\rho,0)$} as a function of isospin asymmetry $X$ at different densities is shown in Fig. 2. Panel (a) of this figure is devoted to results obtained by using only two-body interaction while the other panel shows the results when the {\ph 3BF} is included. In both cases, it is seen that the parabolic law is truly valid in the whole range of $X$ at least up to moderate densities. Therefore, to a good approximation, $E_{sym}(\rho)$ can be determined by using the parabolic law.

Density dependence of the nuclear symmetry energy from both 2BF and 2BF+{\ph 3BF} potentials is shown in Fig. 3. As can be seen, inclusion of {\ph 3BF} causes a significant difference in both the behaviour and the value of $E_{sym}(\rho)$ compared to the case of using only 2BF. The latter leads to a soft symmetry energy that tends to be saturated at high densities, whereas the density-dependent nature of the 3BF provides more repulsive contribution to the PNM EOS and consequently, causes the symmetry energy to increase sharply {\ph with}  density. Exact values of symmetry energy calculated from Eq. (18) are also plotted in this figure. It is seen that there is no significant difference between the exact curves and those obtained from {\ph the parabolic  approximation}, i.e. Eq. (19), as is concluded from Fig. 2.

On the other hand, from Table I one can conclude that the value of nuclear symmetry energy at normal density, $E_{sym}(\rho_0)$, turns out to be too large in the case of using only 2BF, although calculations predict the value of 74.07 MeV for the slope parameter at saturation density which is consistent with the constraint of $ 58.9\pm 16.0 $ MeV \cite{40}. In order to see whether the model predicts the correct curvature parameter, $K_{sym}$, one should use the approximate relation of $K_{asy}\approx K_{sym}-6L$. $K_{asy}$ is the isospin-dependent part of the incompressibility of SNM at normal density (see Eq. (23)) and can be extracted experimentally. Estimate of $K_{asy}=-500\pm50$ MeV has been obtained from isospin diffusion {\ph data} \cite{66,67}, while a systematic study of giant monopole resonance for Sn isotopes predicts $K_{asy}=-550\pm100$ MeV \cite{68}.
The value of this quantity obtained within the LOCV model using 2BF is -520.57 MeV which lies well in the experimentally determined range.  This result shows that the obtained $K_{sym}$ is also reasonable. For the sake of comparison, results from a number of many-body techniques and interactions are also reported in Table I.

Although considering {\ph 3BF} in the LOCV formalism was successful in producing the empirical saturation point, unfortunately it fails to provide a correct value for symmetry energy at normal density. This model predicts the value of $E_{sym}(\rho_0)=39.9$ MeV which although is smaller than the result obtained by using 2BF, is still clearly {\ph too large} compared to the experimentally {\ph determined} range of $\sim 31.6\pm 2.66 $ MeV \cite{40}. Moreover, as can be seen in Fig. 1, the large contribution that comes from the PNM energy, provides a stiff nuclear {\ph symmetry energy} with large slope parameter $L=115.75$ MeV. Obtained value for  $K_{asy}$ is -678.92 MeV which is smaller than the lower limit of -650 MeV predicted from giant monopole resonance for Sn isotopes.

\begin{table}
\caption{\label{tab:table2}%
Symmetry energy at saturation density $E_{sym}(\rho_0)$, slope parameter $L$, curvature parameter $K_{sym}$, and the isospin-dependent part of the isobaric incompressibility $K_{asy}$ (all in MeV) for different parameters of 3BF.
}
\begin{ruledtabular}
\begin{tabular}{lcccc}
\textrm{$(A',U')$}&
\textrm{$E_{sym}(\rho_0)$ }&
\textrm{$L $}&
\textrm{$K_{sym}$}&
\textrm{$K_{asy}$}\\
\colrule
(0,0) &  39.9 & 115.75 & 15.58 & -678.92   \\
(0.0015,-0.000523) D &  35.0 & 89.91 & -47.91 & -587.37   \\
(0.0175,-0.000261) C &  33.0 & 74.04 & -101.42 & -545.66   \\
(0.0337,0) B &  31.0 & 57.43 & -159.13 & -503.71   \\
\end{tabular}
\end{ruledtabular}
\end{table}

\subsection{Revised version of the UIX three-body force with isospin-dependent strengths}
The earliest model of P-wave two-pion exchange potential is due to Fujita and Miyazawa \cite{69}, who assumed that it is entirely due to the excitation of the $\Delta$-resonance. Neglecting the nucleon and $\Delta$ kinetic energies we obtain
 \begin{equation}
A= -\frac{2}{81}\frac{f^{2}_{\pi NN}}{4\pi}\frac{f^{2}_{\pi N \Delta}}{4 \pi}\frac{m^2_\pi}{(m_ \Delta-m_N)},
\end{equation}
where $m_i$ is the mass of particle $i$ and $f_{\pi NN}$ and $f_{\pi N \Delta}$ are the $\pi NN$ and  $\pi N \Delta$ coupling constants, respectively. Using the observed values of $m_\Delta$, $f^{2}_{\pi N \Delta}\sim 0.3-0.35$, and $f^{2}_{\pi NN}\sim0.075$, Eq.(26) predicts that $A\sim -0.04$ MeV \cite{70}.
 In some models of $V_{123}$, by starting from the observed pion-nucleon scattering amplitude, and using current algebra and PCAC constraints, or chiral symmetry, higher absolute value of $A$ is reached which is a consequence of taking into account the contributions of all the $\pi-N$ resonances, as well as that of $\pi-N$ S-wave scattering to the two-pion exchange 3BF. For example, Tucson-Melbourne \cite{71}  and Texas \cite{72}  models predict values of -0.063 MeV and -0.09 MeV for the strength $A$, respectively \cite{70}.
 On the other hand, in all the Urbana models of 3BF,  $A$ is treated as an adjustable parameter and is varied to fit the data.  In light nuclei calculations, parameters  $A$ and  $U$ are chosen to yield the observed binding energies of $^3H$ and $^4He$. On the other, because of the likely many-body effects which are not yet known well,  there is no reason to believe that 3BF in nuclear matter system be the same as in light nuclei. 
As a consequence, in the case of nuclear matter, the values of 3BF parameters are adjusted to yield  the empirical saturation point of SNM as well as the isobaric incompressibility \cite{73,74}.

Because of our poor knowledge of Hamiltonian in high density nuclear systems, there is no reason to believe that the 3BF among nnn be the same as among nnp, npp and ppp.   Therefore, it seems reasonable to correct the Hamiltonian by considering an isospin asymmetry-dependent parametrization for the strength of 3BF.

As already concluded from previous subsection, within the LOCV method, it seems that the current form of {\ph 3BF} fails to reproduce  {\ph the semiempirical values of all} the quantities that characterize the saturation point, at the same time. This fact motivated us to revise the procedure of calculating {\ph 3BF}. Parameters $A$ and $U$ in the UIX three-body force (see Eqs. (12) and (13)) are {\ph usually}  considered as constants which are determined by fitting the empirical saturation density and energy of the cold SNM. In the present work,  we extend this 
procedure, by  proposing an isospin dependent structure for these parameters. {\ph Namely, we assume } that $A$ and $U$ {\ph can  be} expanded to second-order in asymmetry parameter $X$ as
\begin{equation}
A=A_{sym}+A'X^2,
\end{equation}
and
\begin{equation}
U=U_{sym}+U'X^2.
\end{equation}

\gii It must be noted that this consideration will not violate the formalism of LOCV method, since one still has two, but different parameters in 3BF for each given nuclear system i.e., SNM, PNM  and $\beta$-stable matter. Rewriting the 3BF strengths in the form of Eqs. (27) and (28) only changes the nuclear Hamiltonian  and it surely keeps the many-body formalism unchanged.
  Such correction in Hamiltonian is also seen in other papers such as that by Akmal, et al \cite{75}, where in addition to 2BF and 3BF, the authors consider a density-dependent term with adjustable strength in order to reproduce correctly the saturation properties of SNM.
  
By putting $X=0$  in Eqs. (27) and (28), i.e. considering symmetric nuclear matter, one reaches the previous scenario. Therefore, $A_{sym}$ and $U_{sym}$ must have the same values as those mentioned before, i.e. -0.041 MeV and 0.000523 MeV, respectively. Variables $A'$ and $U'$ are determined in such a way that both the symmetry energy and its slope lie in the empirical range.
For the whole range of $A'$ and $U'$ reported in Fig.4, values of both  $A$ and $U$ lie well in the range of $-0.09<A<-0.0058$ MeV \cite{72,76} and $0<U<0.032$ MeV \cite{77,70} which are obtained by other authors, in case of any nuclear matter system (SNM, PNM and $\beta$-stable matter).

\begin{table}
\caption{\label{tab:table2}%
Kinetic and potential energy contributions to 
 $E_{sym}$, $L$ and $K_{sym}$  (all in MeV) at saturation density for different parameters of 3BF.}
 
\begin{ruledtabular}
\begin{tabular}{lcccc}
\textrm{$(A',U')$}&
\textrm{ }&
\textrm{$E_{sym}$}&
\multicolumn{1}{c}{\textrm{$L$}}&
\textrm{$K_{sym}$}\\
\colrule
(0,0) &$\langle T \rangle$ & 13.75 & 27.49 & -27.50 \\
      &$\langle V \rangle$ & 26.15 &88.26 & 43.08 \\
      &Total& 39.90 & 115.75 & 15.58 \\  
(0.0015,-0.000523) D &$\langle T \rangle$ & 13.75 & 27.49 & -27.50 \\
                   &$\langle V \rangle$& 21.25 & 62.42  & -20.41 \\
                   &Total& 35.00 & 89.91 & -47.91  \\
(0.0175,-0.000261) C &$\langle T \rangle$ & 13.75 & 27.49 & -27.50  \\
                    &$\langle V \rangle$ & 19.25 & 46.55 & -73.92 \\ 
  			        &Total & 33.00 & 74.04 & -101.42  \\
(0.0337,0) B  &$\langle T \rangle$ & 13.75 & 27.49 & -27.50 \\ 
              &$\langle V \rangle$ & 17.25 & 29.94  & -131.63 \\       
              &Total & 31.00 & 57.43 & -159.13  \\ 
\end{tabular}
\end{ruledtabular}
\end{table}

In order to achieve the proper value for  $E_{sym}(\rho_0)$  and $L$, a softer EOS for PNM should be produced. Therefore, the repulsive part of {\ph 3BF, given by}  Eq. (13), which is controlled by the strength $U$, should be smaller. One way to do this task is to assume that there is no repulsive term in {\ph 3BF} for PNM $(X=1)$. This {\ph aim} can be reached by putting $U'=-U_{sym}$ in Eq. (28). Another way is to assume that the repulsive part of {\ph 3BF} for PNM is the same as that for {\ph SNM} which means setting $U'=0$ {\ph in Eq. (28)}. Choosing positive values for  $U'$ will increase the repulsive component and does not help solving the problem. On the other hand, if one {\ph sets}  $U'<-U_{sym}$, {\ph expression given by} Eq. (28) becomes negative for PNM which is against the nature of this phenomenological repulsive component. As a consequence, $U'$ can vary between two extreme values mentioned above, i.e., $-U_{sym}<U'<0$.

 {\ph Figure 4} shows the resulting $E_{sym}(\rho_0)$  and $L$  in the plane of $A'-U'$. Each shaded parallelogram shows a specific span for symmetry energy at normal density while each diagonal black  line corresponds to a fixed value of slope parameter. As an example, all pairs $(A',U')$ which are located within the {\ph tilted} box produce $31<E_{sym}(\rho_0)<32$ MeV. Similarly, all pairs $(A',U')$ which are between the lines $L=60$  and $L=65$, produce $60<L<65$ MeV. As can be seen in this figure, by setting $A'=0$ in Eq. (27) one reaches to $E_{sym}(\rho_0)>35$ MeV and  $L>90$ MeV for the whole range of  $U'$ which are large compared to empirical values. Therefore, $A'$ should be also non-zero in order to get smaller symmetry energy as well as the slope parameter. From this figure, it can also be concluded that both $E_{sym}(\rho_0)$ and $L$ vary with parameters $A'$ and $U'$, although the slope parameter seems to be more sensitive. Particularly, for a fixed value of $U'$, both the symmetry energy and the slope parameter decrease with $A'$  while at fixed  $A'$ both $E_{sym}(\rho_0)$ and $L$ increase as $U'$ tends to zero.

In order to understand the difference between the revised version of 3BF (which is characterized by Eqs. (27) and (28)) and the old one (the case of $A'=U'=0$), a comparison is made for each component of the effective two-body interaction, namely the coefficients of Eq. (16). To do this task, three $(A',U')$ pairs which are denoted by letters B, C, and D in Fig. 4,  are compared with the {\ph 3BF} with isospin-independent strengths, i.e. the case of $A'=U'=0$. Components $V^{{2\pi}}_{\sigma \tau } (r)$, $V^{{2\pi}}_{t}(r)$, and $V^{{R}}_{c}(r)$ of these four points are respectively plotted in Figs. 5 to 7 at $\rho=0.17~{\ph\rm fm^{-3}}$ for PNM. Corresponding numerical values of $A'$ and $U'$ are also reported in Table II. Two-pion exchange contribution to the UIX 3BF is divided into two components, namely $V^{{2\pi}}_{\sigma \tau } (r)$ and $V^{{2\pi}}_{t}(r)$ which are both controlled by the strength  $A$.
 Now let us consider the PNM case which is reached by putting $X=1$ in Eqs. (26) and (27){\ph , and keeping  in mind} that $A_{sym}$ is {\ph negative}, as mentioned earlier.
  Consequently, the strength $A$ tends to zero by increasing $A'$ and one expects to get two-pion exchange components with smaller absolute value.
This fact can be {\ph clearly seen }in Figs. 5 and 6. Among four chosen cases, the largest {\ph value of $A'$ corresponds} to point B (see Fig. 4). Therefore, the smallest absolute value of effective two-body potential is obtained in this case.

\begin{table}
\caption{\label{tab:table2}%
Partial wave decomposition of the potential part of 
 $E_{sym}$, $L$ and $K_{sym}$ (all in MeV) at saturation density for different parameters of  3BF.} 
\begin{ruledtabular}
\begin{tabular}{lcccc}
\textrm{Partial wave}&
\textrm{$(A',U')$}&
\textrm{$E_{sym}(\rho_0)$}&
\multicolumn{1}{c}{\textrm{$L$}}&
\textrm{$K_{sym}$}\\
\colrule
$^{1}S_{0}$ & (0,0)  & 0.47 & 11.58 & 4.76 \\
$ $ & (0.0015,-0.000523) D  & -1.27 & 5.29 & 0.45 \\
$ $ & (0.0175,-0.000261) C  & -1.1 & 4.67 & -5.35 \\
$ $ & (0.0337,0) B & -0.97  & 3.76 & -12.04 \\

$^{3}S_{1}$ & (0,0)  & 25.87  & 61.76 & -29.41 \\
$ $ &  D  &    &   &  \\
$ $ &  C  &    &   &   \\
$ $ &  B  &    &   &   \\

$^{1}P_{1}$ & (0,0)  & -4.67  & -19.19 & -26.36 \\ 
$ $ &  D  &    &   &   \\ 
$ $ &  C  &    &   &   \\ 
$ $ &  B  &    &   &   \\ 

$^{3}P_{0}$ & (0,0)  & -0.76 & 0.08 & 8.4\\
$ $ &  D  & -1.03 & -1.38 & 4.92\\
$ $ &  C  & -0.73 & 0.96 & 14.1\\
$ $ &  B & -0.44  & 3.19 & 21.17\\

$^{3}P_{1}$ & (0,0) & 4.38  & 19.21 & 40.08\\
$ $ &  D  & 3.77 & 14.74 & 23.97\\
$ $ &  C  & 3.64 & 13.13 & 15.08\\
$ $ &  B & 3.59  & 11.64 & 5.99\\

$^{3}P_{2}$ & (0,0)  & 0.03 & 12.25 & 54.53 \\
$ $ &  D  & -1.86 & 1.18 & 23.80 \\
$ $ &  C  & -3.63 & -13.09 & -30.63 \\
$ $ &  B & -5.43  & -27.83 & -86.59 \\

$^{1}D_{2}$ & (0,0)  & -2.50 & -10.00 & -4.87\\
$ $ &  D  & -2.76 & -11.78 & -11.72\\
$ $ &  C  & -2.43 & -9.19 & -1.71\\
$ $ &  B & -2.10  & -6.56 & 8.60\\

$^{3}D_{1}$ & (0,0)  & -0.14  & -0.64 & -7.33 \\ 
$ $ &  D  &   &   &  \\
$ $ &  C  &   &   &   \\
$ $ &  B  &   &   &   \\

$^{3}D_{2}$ & (0,0)  & 3.37& 13.83 & 14.86 \\
$ $ &  D  &  &   &   \\
$ $ &  C  &  &   &   \\
$ $ &  B  &  &   &   \\

$^{3}F_{2}$ & (0,0)  & 0.09 & -0.61 & -11.58 \\ 
$ $ &  D  & -0.02 & -1.39 & -13.59 \\ 
$ $ &  C & -0.91 & -5.69 & -17.17 \\ 
$ $ &  B  & -1.82 & -10.02 & -20.52 \\ 

\end{tabular}
\end{ruledtabular}
\end{table}

Figure 7 shows the repulsive component of UIX {\ph 3BF} whose strength is governed by parameter $U$. Once again, the isospin asymmetry is set to 1 {\ph (the case of PNM)}. As already {\ph pointed out},  $U_{sym}$ is a positive {\ph parameter} whereas $U'$ is a negative one which varies in the range of $-U_{sym}<U'<0$. Therefore, {\ph larger} negative values of $U'$ lead to {\ph smaller $U$} and consequently weaker repulsive components, as is seen in Fig.7. On the other hand, point D corresponds to the lower limit predicted for $U'$,  i.e. $U'=-U_{sym}$ which is equivalent to the case of $U=0$. Therefore, {\ph at point D there is } no repulsive component.

In Fig. 8(a) the energy per nucleon of PNM is presented which is obtained by using four different effective two-body potentials plotted in Figs. 5 to 7. It is seen that the EOS is strongly sensitive to the proposed parametrization of variables $A$ and $U$. All curves with non-zero $A'$ and $U'$ are located below the case of $A'=U'=0$ at the whole range of density. Therefore, it can be concluded that by producing an isospin-dependent structure for {\ph the  3BF parameters}, our wish for gaining a softer EOS for PNM can be fulfilled.

Figure 8(b) represents the corresponding results for the density dependence of nuclear symmetry energy. A significant difference can be seen in both the value and behaviour of $E_{sym}(\rho)$ depending on the values of $A'$ and $U'$. While symmetry energy obtained from  $A'=U'=0$ parametrization increases with density, those of isospin-dependent $A$ and $U$ tend to saturate at large densities. This tendency decreases from point B to D. Values of symmetry energy at saturation density $E_{sym}(\rho_0)$, slope parameter $L$, curvature parameter $K_{sym}$, and the isospin-dependent part of the isobaric incompressibility $K_{asy}$ are reported in Table II for all curves. From this Table it is clear that all mentioned quantities which characterize the empirical saturation point, lie well within the experimentally predicted ranges.

 In Fig. 8(c) the density dependence of $E_{sym}$ predicted by LOCV method  {\phh assuming model 
D of revised UIX 3BF (curve D in Fig. 8(b))} is compared with the results of other many-body approaches such as variational \cite{31} and BHF \cite{69,70,71} models using different potentials. It is seen that both the values and behaviour of $E_{sym}$ {\phh obtained using D version of  revised UIX 3BF  in the LOCV method is in agreement with results of BHF calculations with 3BF. }

One of the quantities which is directly affected by the symmetry energy is the relative population of protons 
{\phh  $Y_p=\frac{1}{2}(1-X)$} in the $\beta$-stable matter which is a matter composed
of an uncharged mixture of neutrons, protons, electrons, and
muons in equilibrium with respect to the weak interaction. Figure 9 displays this quantity as a function of total baryon density for different  $(A',U')$ pairs, namely $(0,0)$, points B to D and $(0.0127,-0.000523)$, denoted by letter E. It is seen that both the value and behaviour of the proton fraction is strongly sensitive to the values of 3BF strengths. some curves are rising in the whole baryon density range while others tend to saturate. The reason is related to the behaviour of the symmetry energy  in each case. As is seen in Fig. 8(b), $E_{sym}(\rho)$ saturates for some values of $A'$ and $U'$ and consequently, the proton fraction shows the same 
{\ph behaviour, due to {\phh $Y_p(\rho)\propto E^3_{sym}(\rho)$, relation valid for $Y_p\ll 1$ which is satisfied 
here}  (see, e.g. \cite{72}). } 
The {\ph smaller} are the symmetry energy and the slope parameter, the smaller is  the density {\ph at which the proton fraction saturates}. However, there exists a region in $(A',U')$ plane characterized by $-0.000327<U'<0$ and $E_{sym}(\rho_0)\gtrsim 33.6$ MeV (see Fig. 4) {\ph within which} proton fraction increases monotonously and does not saturate in the whole range of density up to {\ph $1~{\rm fm^{-3}}$} if one chooses any $(A',U')$ pairs from this area to calculate the energy of  $\beta$-stable matter.


Since three-body forces  which are one of the key {\ph quantities} in calculations of the symmetry energy,  {\ph contribute only} to the potential part of $E_{sym}(\rho)$, it seems valuable to separate the symmetry energy into its kinetic, $\langle T \rangle$, and potential, $\langle V \rangle$, parts. Results of such separation is reported in Table III for different $(A',U')$ pairs at saturation density $\rho_0=0.1748 {\ph {~\rm fm^{-3}}}$. The kinetic term of the symmetry energy is,   to a very good approximation,  equal to the difference in the kinetic energy between PNM and SNM, calculated using a simple Fermi gas model. Therefore, this term gives the same contribution {\ph for all} parametrizations of  3BF. In the case of $A'=U'=0$, the large contribution comes from the potential {\ph part, and this  causes} both the symmetry energy and its slope not to lie in the {\ph experimental}  range. This problem is resolved {\ph for parametrizations B,C,D } by considering isospin-dependent parametrization for the {3BF coefficients}.   As can be concluded in Fig. 4, the least value of $\langle V \rangle$ {\ph corresponds}  to point B.

In the LOCV approach one has access to the separate contributions of partial waves energies in the correlated many-body state. Table IV shows the partial wave decomposition of the potential part of  $E_{sym}(\rho_0)$, $L$, and $K_{sym}$ up to $J=2$ for different parameters of {\ph 3BF}. Note that a number of {\ph partial waves}, namely $^{3}S_{1}$, $^{1}P_{1}$, $^{3}D_{1}$, and $^{3}D_{2}$, does not contribute to neutron matter. Therefore,  contributions from these waves to the symmetry energy are the same for all allowed  values of $A'$ and $U'$. {\ph For} the six remaining {\ph partial waves}, all quantities $E_{sym}(\rho_0)$, $L$, and $K_{sym}$ decrease (or increase) monotonically from point D to B. The only exception is $^{1}S_{0}$ channel in which $E_{sym}(\rho_0)$ increases from D to B while $L$ and $K_{sym}$ show opposite behaviour. It can be concluded from this table that the $^{3}S_{1}-^{3}D_{1}$ channel gives the major contribution to  $E_{sym}(\rho_0)$ and $L$. It is also observed  that the $S=1$ (spin triplet) and  $T=0$ (isospin singlet) channel gives the largest contribution to both symmetry energy and its slope for all values of $A'$ and $U'$. Similar {\ph features} have been reported in \cite{7} and references therein.
{\ph 
\subsection{Some astrophysical implications of our results}
We expect that for $0.5\rho_0<\rho <{\phh 2\rho_0})$ ({\it outer core}, see e.g. \cite{72}) NS matter is an uniform, electrically neutral $npe\mu$ plasma in $\beta$-equilibrium. This is also the simplest (minimal) model of the {\it inner core} of NS ($\rho>2\rho_0$). Using our 2BF+3BF nuclear Hamiltonian, we calculated    the equation of state (EOS)  and composition of  NS core.  Beta equilibrium implies specific fractions of $j=n,p,e, \mu$ in  NS matter constituents, $Y_j=\rho_j/\rho$.  

Cooling rate of NS core dramatically depends on  $Y_p$. Namely, if $Y_p<Y_{\rm DU}$, where $Y_{\rm DU}= 0.11 - 0.145$ (depending on the precentage of muons), then powerful direct Urca (Durca)  process of neutrino emission can act, leading to a fast NS cooling  \cite{73,74}(for review of neutrino cooling of NS see, e.g. \cite{75}). As we see in Fig.9, in the case of our 2BF+3BF Hamiltionian, 3BF decides whether the Durca process is acting in NS core (this was already remarked in \cite{76}). 

Making 3BF less repulsive is crucial  for getting correct values of $E_{sym}(\rho_0)$ and $L$.  However, this leads to a softer EOS for NS cores. As a consequence, the maximum allowable mass for NS for the 2BF+3BF Hamiltonian, $M_{\rm max}$, decreases (in what follows we neglect the effect of rotation on $M_{\rm max}$, which for the frequencies of recently discovered $2\;M_\odot$ pulsars \cite{73,74} is very small). For all models of our revised UIX 3BF we get $M_{\rm max}>2\;M_\odot$, consistently with \cite{77,78}. 

This subsection is but a brief report on the effects of a revised 3BF on NS models. More detailed presentation will be given in a separate paper.

}
\section{CONCLUSION}

The extended LOCV formalism including the UIX {\ph 3BF} force is used in order to study the density dependence of nuclear symmetry energy at zero temperature. The AV18 potential is {\ph used} as the bare two-body interaction. It is shown that although inclusion of {\ph 3BF} is successful in reproducing the saturation point properties, namely saturation density and saturation energy as well as the isobaric incompressibility, it fails to predict a reasonable value for both symmetry energy at normal density and its slope. In order to {\ph correct this deficiency}, a revised version of {\ph 3BF} is proposed which is based on an isospin-dependent parametrization of {\ph coefficients}  $A$ and $U$ in the UIX {\ph  model}.  {\ph To analyse the effect} of the new parametrization, different components of the revised {\ph 3BF are}  compared to those of the old one. Effect of the new parametrization of {\ph 3BF} on density dependence of symmetry energy, its value at saturation density and its slope and curvature is also studied.  It is  shown that using this kind of {\ph 3BF} in the LOCV formalism results in {\ph acceptable} values for  $E_{sym}(\rho_0)$, $L$, $K_{sym}$ and $K_{asy}$. Moreover, separate contributions of partial waves energies in the potential part of symmetry energy {\ph are} reported. It is {\ph found}  that the  $S=1$ and  $T=0$ channel, and particularly the $^{3}S_{1}$ {\ph partial} wave, give the largest contribution to both symmetry energy and its slope for all values of $A'$ and $U'$.

{\ph Implications of our 3BF on NS models  were briefly mentioned in Sect. IV.C. A detailed report on the application of our 3BF models to modeling of NS structure and evolution is being prepared.}

\begin{acknowledgments}
HRM thank the Research Council of University of Tehran for the grants. 
{\ph This work was supported in
part by the National Science Centre, Poland, under the grant No. {\phh 2014/13/B/ST9/02621}.}
\end{acknowledgments}


\begin{figure*}
\includegraphics[scale=0.5]{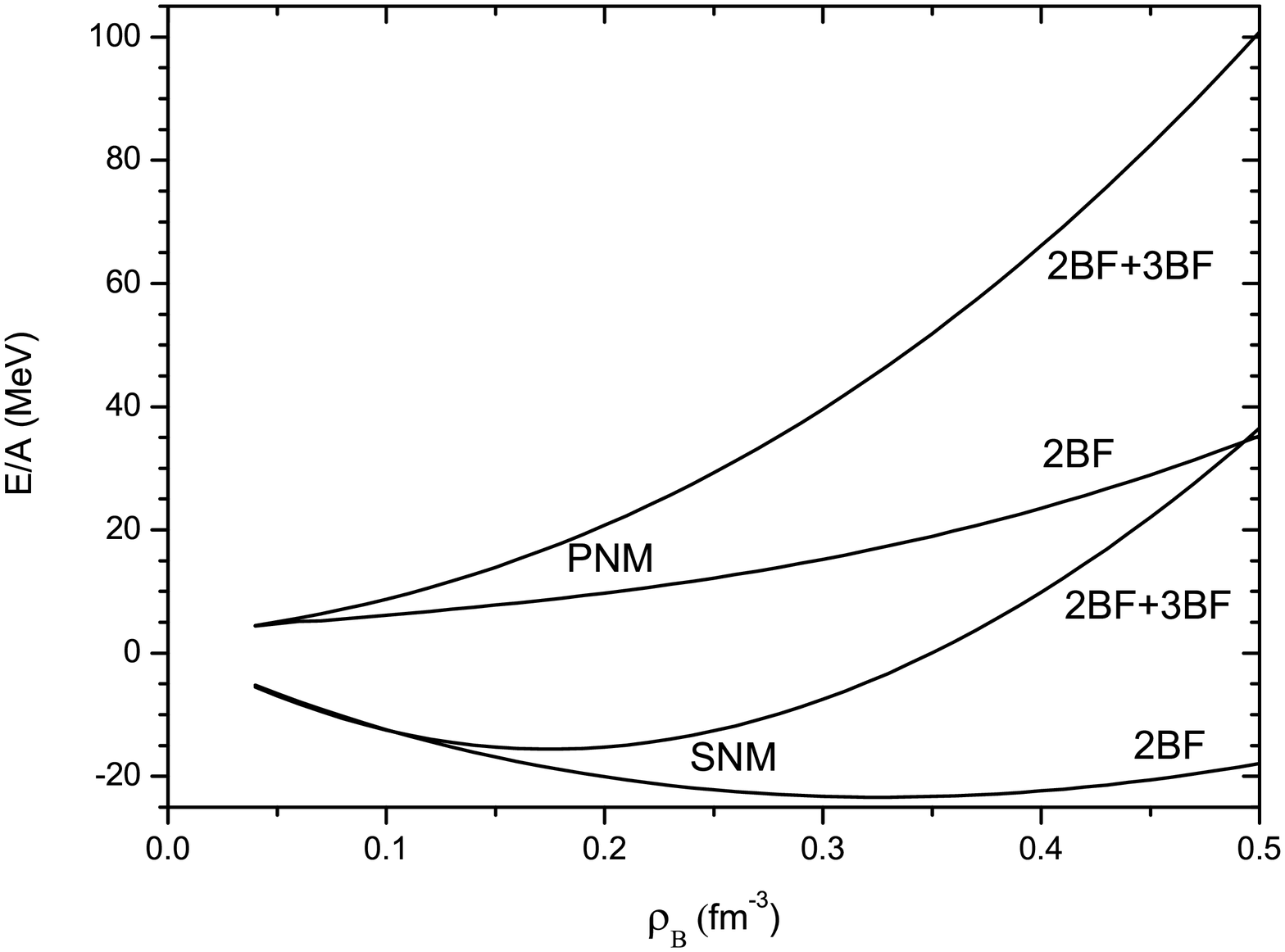} 
\caption{\ph Equation of state of SNM $(X=0)$ as well as PNM $(X=1)$ using different 2BF and 2BF+3BF.}\label{fig 1.}
\end{figure*}

\begin{figure*}
\includegraphics[scale=0.5]{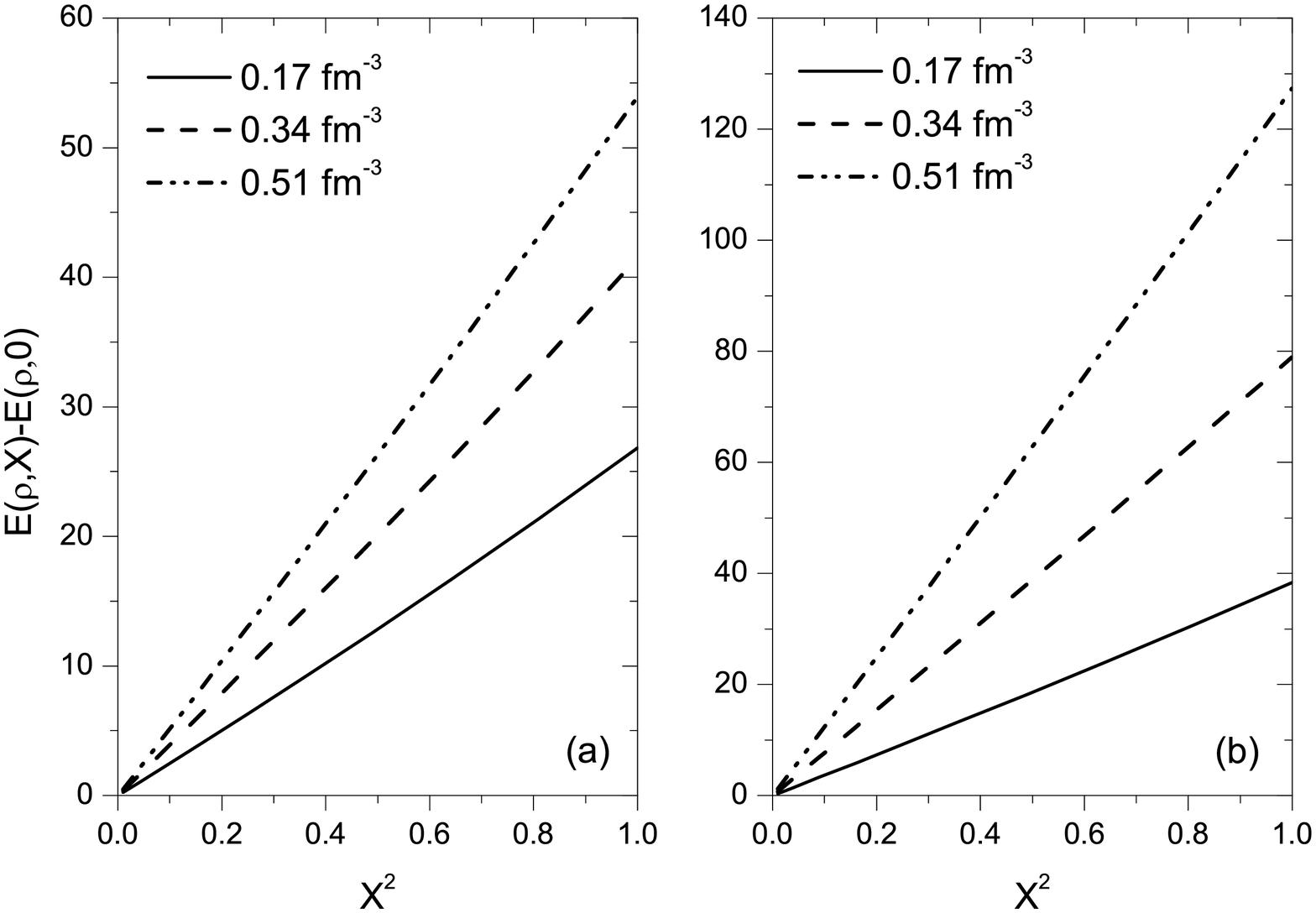} 
\caption{\ph (a) Quadratic dependence of $E(\rho,X)-E(\rho,0)$ as a function of asymmetry parameter using 2BF for different  densities. (b) Same as (a) but for 2BF+3BF.}\label{fig 2.}
\end{figure*}

\begin{figure*}
\includegraphics[scale=0.5]{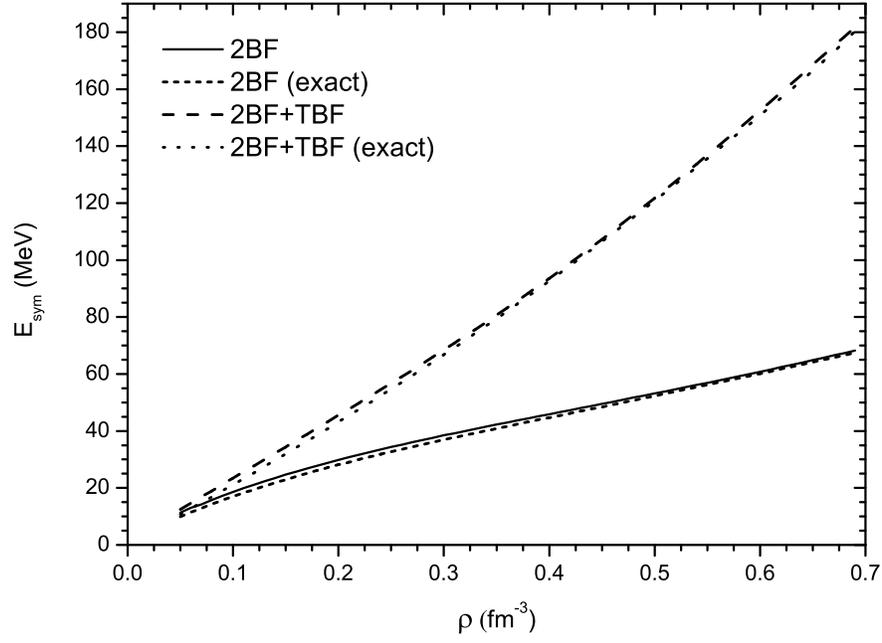} 
\caption{Density dependence of nuclear symmetry energy  for different  interactions. }\label{fig 3.}
\end{figure*}

\begin{figure*}
\includegraphics[scale=0.5]{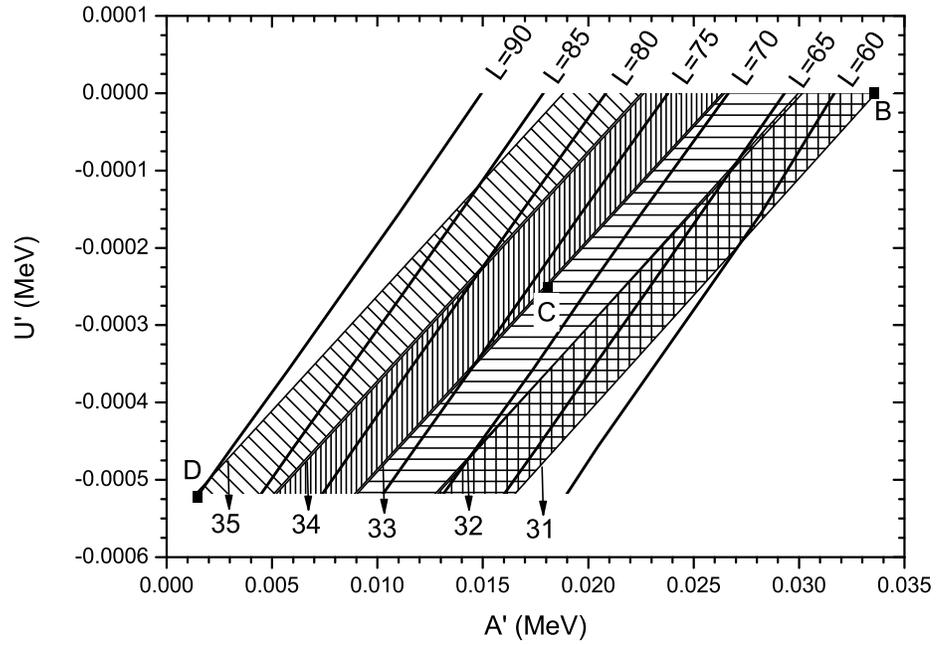} 
\caption{Symmetry energy at saturation density $E_{sym}(\rho_0)$ as well as its slope $L$ (all in MeV) in the plane of $A'-U'$. Numerals 31 to 35 {\ph denote} the value of $E_{sym}(\rho_0)$. See the text for details.}\label{fig 4.}
\end{figure*}

\begin{figure*}
\includegraphics[scale=0.5]{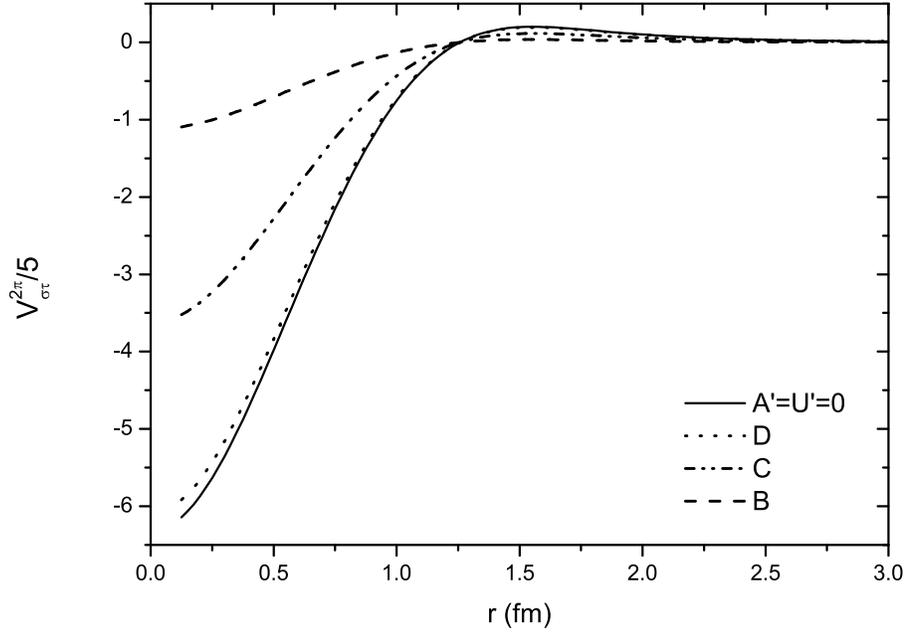} 
\caption{The $V^{{2\pi}}_{\sigma \tau } (r)$ {\ph component} of the effective two-body potential at $\rho=0.17{\ph \;{\rm fm^{-3}}}$ for PNM obtained by using different parameters of 3BF. Points B to D are {\ph indicated}  in Fig. 4. }\label{fig 5.}
\end{figure*}

\begin{figure*}
\includegraphics[scale=0.5]{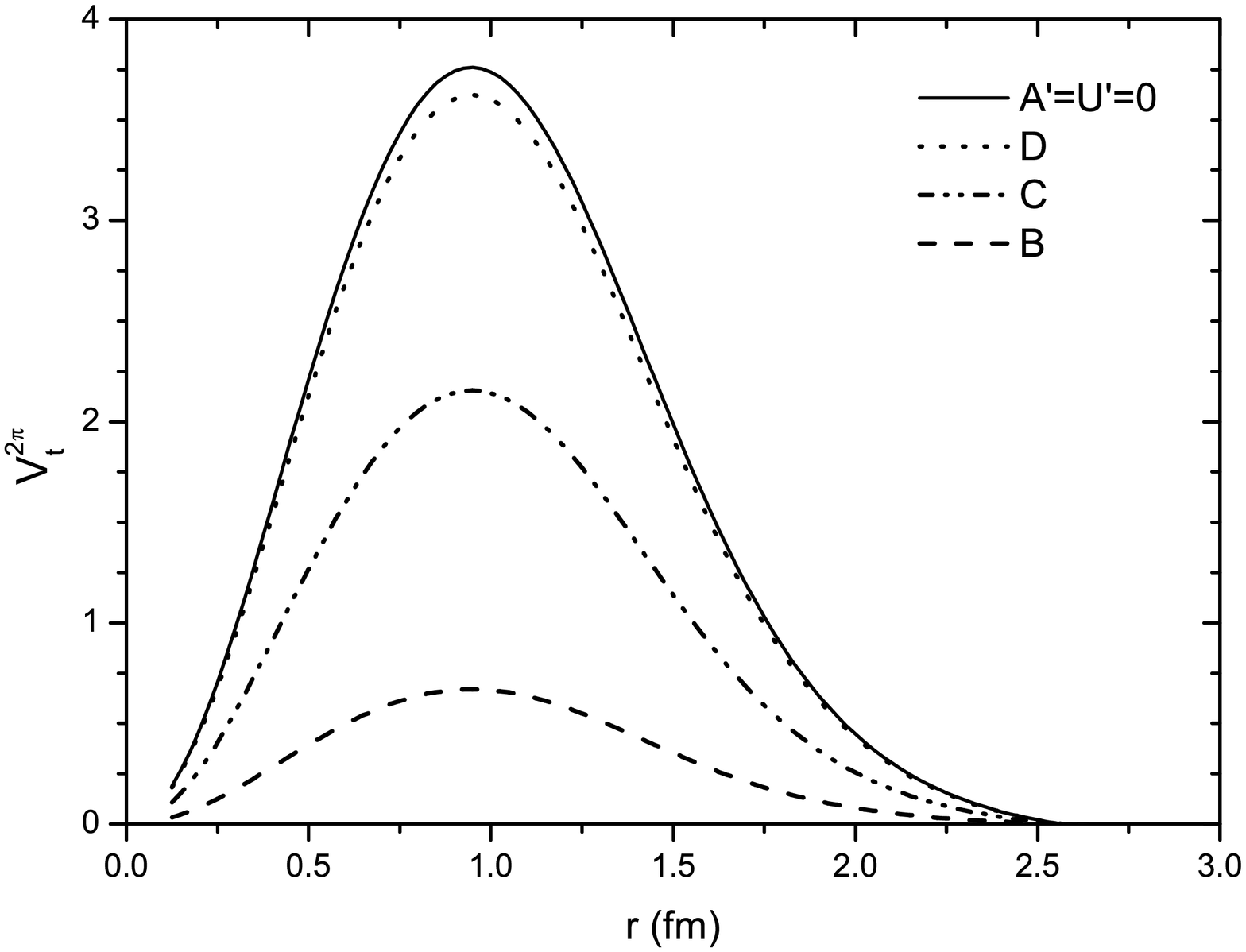} 
\caption{Same as Fig. 5 but for $V^{{2\pi}}_{t } (r)$ component. }\label{fig 6.}
\end{figure*}

\begin{figure*}
\includegraphics[scale=0.5]{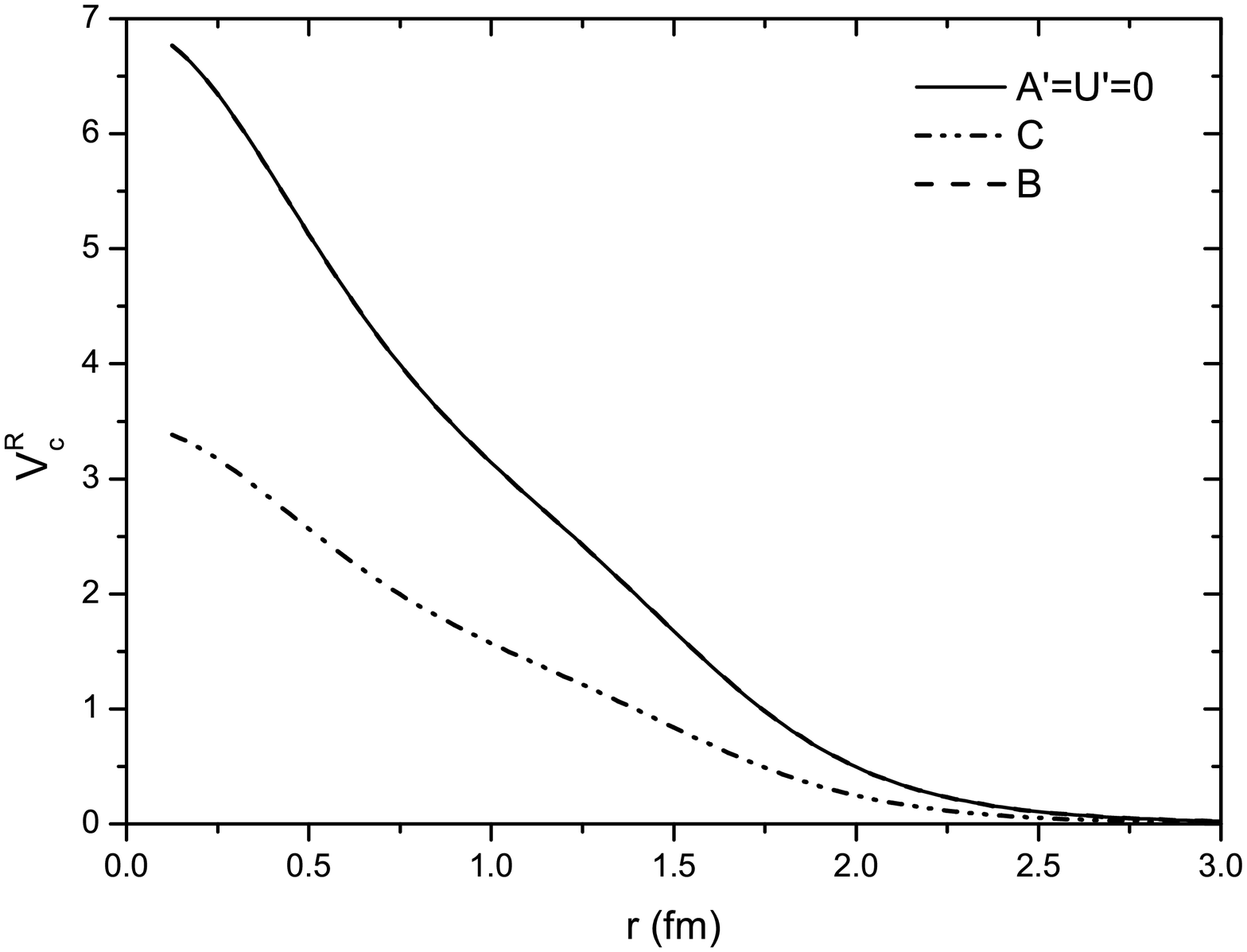} 
\caption{Same as Fig. 5 but for $V^{R}_{c } (r)$ component. }\label{fig 7.}
\end{figure*}

\begin{figure*}
\includegraphics[scale=0.5]{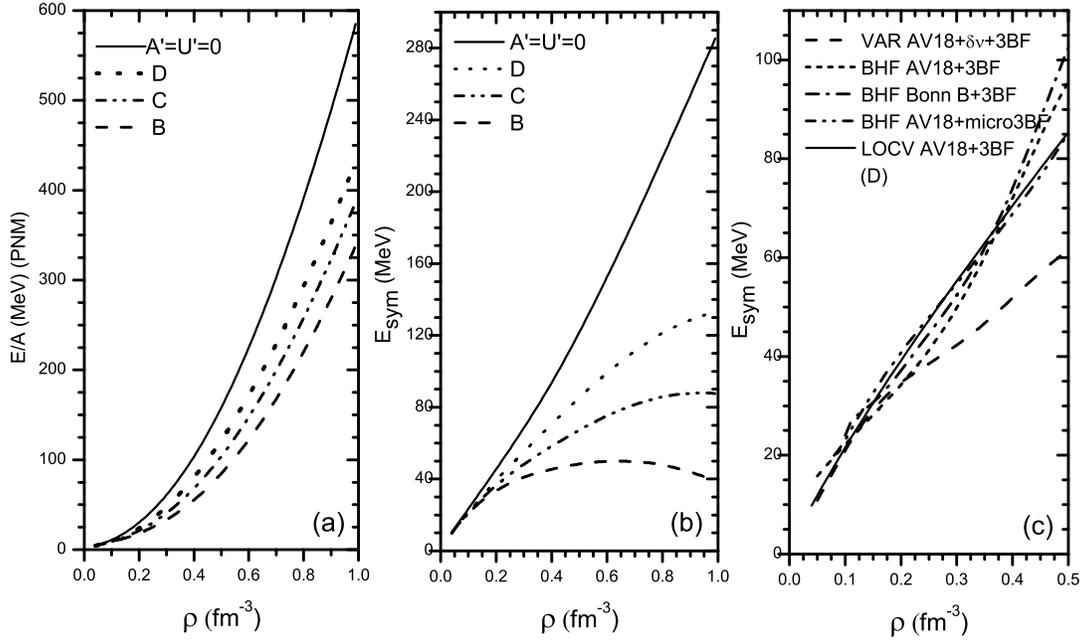} 
\caption{(a) {\ph Energy} per nucleon of PNM for different revised 3BF parameters. (b) Same as panel (a) but for nuclear symmetry energy.{\gi (c) Density dependence of nuclear symmetry energy obtained by LOCV approach using AV18+3BF (curve D in panel (b)) as well as other many-body techniques and potentials.} }\label{fig 8.}
\end{figure*}

\begin{figure*}
\includegraphics[scale=0.5]{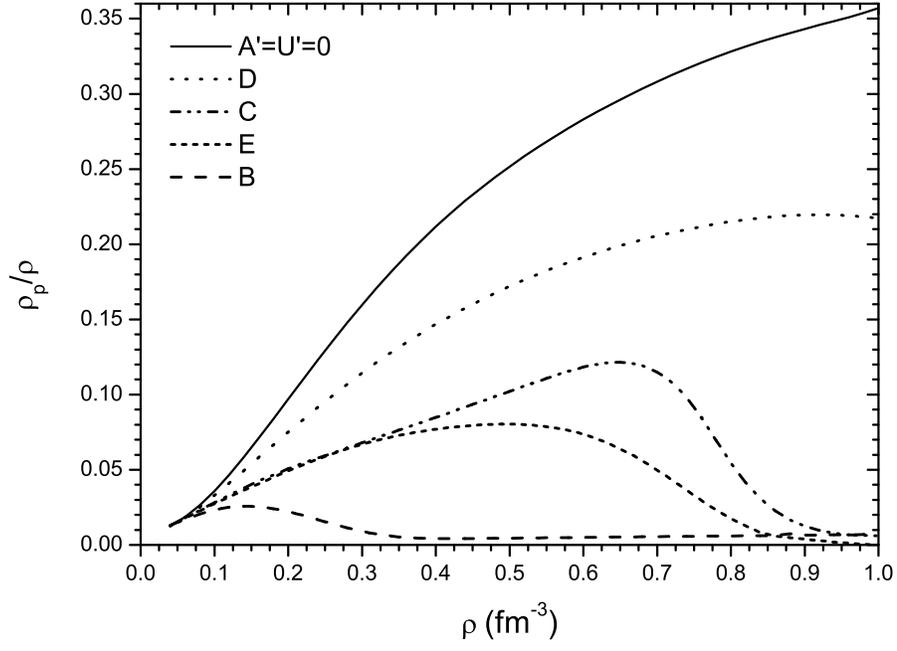} 
\caption{\ph Fraction  of protons in $\beta$-stable matter for different revised 3BF parameters.  }\label{fig 9.}
\end{figure*}

\end{document}